\documentclass[pra, 10pt, twocolumn, letter, floatfix]{revtex4}

\usepackage{amsmath}
\usepackage{amsfonts}
\usepackage{amssymb}
\usepackage{graphicx}
\usepackage{dcolumn}
\usepackage{bm}
\usepackage{subfig}

\begin{document}

\title[pADCCD] {Amplitude determinant coupled cluster with pairwise doubles}

\author{Luning Zhao$^{1}$}
\author{Eric Neuscamman$^{1,2,}$\footnote[1]{Electronic mail: eneuscamman@berkeley.edu}}
\affiliation{$^1$Department of Chemistry, University of California, Berkeley, California 94720, USA\\
             $^2$Chemical Sciences Division, Lawrence Berkeley National Laboratory, Berkeley, California 94720, USA}

\date{\today}


\begin{abstract}
Recently developed pair coupled cluster doubles (pCCD) theory successfully reproduces doubly occupied configuration interaction (DOCI) with mean field cost. 
However, the projective nature of pCCD makes the method non-variational and thus hard to improve systematically.
As a variational alternative, we explore the idea of coupled-cluster-like expansions based on amplitude determinants
and develop a specific theory similar to pCCD based on determinants of pairwise doubles.
The new ansatz admits a variational treatment through Monte Carlo methods while remaining size-consistent and, crucially, polynomial cost.
In the dissociations of LiH, HF, H$_2$O and N$_2$, the method performs very similarly to pCCD and DOCI, suggesting that coupled-cluster-like ansatzes
and variational evaluation may not be mutually exclusive.
\end{abstract}

\maketitle


\section{Introduction}
\label{sec:introduction}

The electronic correlation energy can be divided into two parts\cite{MolElecStruc}. Dynamic correlation, small changes to mean field theory, can be described 
effectively through many-body perturbation theory\cite{Plesset:1934:mp2}. Static or non-dynamic correlation becomes important when the ground state contains degenerate 
or near-degenerate configurations, i.e. where more than one determinant is needed for a qualitative description of the system. 
Such situations arise in bond breaking\cite{MolElecStruc}, large $\pi$-conjugated
molecules where the HOMO-LUMO gap is small\cite{Parac:2003:failure_tddft}, transition metal complexes\cite{Eric:2016:ScO} and superconductivity\cite{Subedi:2008:supercond}. 
For such cases, a single Slater determinant gives qualitatively incorrect results. 

The concept of seniority has a long history in nuclear physics\cite{NuclearMB}. Seniority number is the measurement of unpaired electrons in a determinant. 
It has been showed before that strong correlation is very often concentrated within the low seniority regions of Hilbert space\cite{Scuseria:2011:seniority}.
Among them, the seniority zero wave function is the most studied due to the fact that it is both size consistent and capable of capturing a large amount of 
strong correlation. 

The most elaborate seniority zero wave function is doubly occupied configuration interaction (DOCI)\cite{Wilson:1967:doci}, which is a full configuration
interaction (FCI) wave function
limited to seniority zero space. DOCI was thoroughly studied in the pioneering days of quantum chemistry\cite{Wilson:1967:doci, Shull:1962:Be, Fogel:1965:Be} 
and has many desirable features, being size-consistent, variational and effective for describing strong correlation. 
However, DOCI's factorial cost scaling severely limits its applicability. 

Coupled Cluster (CC) with single and double excitations (CCSD)\cite{Barlett:1982:ccsd} and CCSD with perturbative triple excitations [CCSD(T)]\cite{Martin:1989:ccsd(t)} 
provide a very powerful way to describe dynamic correlation for a polynomial scaling cost. Between its rigorous size consistency, its systematic improvability, 
and its exceptional accuracy even at the CCSD(T) level, CC has become one of the most trusted and even routine methods in quantum chemistry for weakly correlated 
systems that are not too large.
Unfortunately, the main drawback of traditional CC is that the ansatz is optimized in a projective manner to achieve polynomial scaling, which makes the energy non-variational
and leads to qualitative failure in strongly correlated regimes \cite{Scuseria:2014:pccd,Scuseria:2014:pccd2,Scuseria:2015:cc_strong_corr}.
Examples include breaking of multiple bonds simultaneously, 
and the Hubbard model at large U, in which RHF-based CCSD or even CCSD(T) ``overcorrelates'' and the converged energy is well below the FCI energy \cite{Barlett:2007:cc_rev}.
While many approaches have been taken to resolve this issue, including multi-reference CC \cite{barlett:2012:mrcc},
combinatorially scaling variational CC\cite{Knowles:2010:vcc}, single-pair couple cluster\cite{Scuseria:2016:ccd0} 
and CC valence bond \cite{Martin:2014:ccvb}, there remain no CC methods that can capture both static and dynamic correlation in large molecules at an affordable cost.

Recently, the antisymmetric-product of one-reference-orbital geminals (AP1roG)\cite{Ayers:2013:ap1rog} was introduced by Ayers and coworkers in an attempt to deal with 
static correlation at polynomial cost. Impressively, they showed that AP1roG gives results almost equivalent to those of DOCI.
Later, Scuseria and coworkers recognized that AP1roG is a simplified version of coupled cluster doubles (CCD), in which electrons are 
paired and only pair excitations are allowed, which they termed pair-CCD (pCCD)\cite{Scuseria:2014:pccd}.
It is quite remarkable that by eliminating the vast bulk of the cluster operator, 
the pCCD ansatz accurately reproduces the factorially complex DOCI. It is also important: since DOCI provides a powerful description of strong correlation in a wide variety of 
systems, so do AP1roG and pCCD. However, like most other CC methods, both pCCD and AP1roG are non-variational. 

One may raise the question that since pCCD is already so close to DOCI, what is the point of achieving variationality? One should note that although pCCD 
does not have enough degrees of freedom to break variationality, once one breaks electron pairs and adds high seniority 
determinants to the ansatz, aiming to recover the dynamic correlations missing in pCCD, this theory fails in the same manner as traditional CC, yielding non-variational 
energies\cite{Scuseria:2014:pccd}. Therefore, it is not trivial to improve pCCD systematically. 

If one optimizes the wave function parameters in coupled cluster ansatz variationally, 
the unphysical overcorrelation in multiple bond breaking can be eliminated\cite{Knowles:2010:vcc}. This variational CC (VCC) approach 
has combinatorial cost scaling, however, and can only be applied to 
small systems. As a compromise, Knowles and Robinson developed quasi-variational coupled cluster theory\cite{Knowles:2012:quasi-vcc}. This method is only approximately
variational, however, and the energies are not strictly upper bounds.

Besides deterministic methods, variational quantum Monte Carlo (VMC)\cite{Umrigar:2015:qmc} is another accurate and versatile method for treating electron correlation. 
The VMC method is used to find expectation values of operators for a given trial wave function and to optimize the parameters in the trial wave function stochastically. 
In general VMC need not depend on the independent particle approximation as it is compatible with more general ansatzes such as 
Jastrow Slater, Jastrow multi-Slater\cite{Umrigar:2015:qmc} and Jastrow AGP\cite{Sorella:2003:jagp}. VMC is strictly variational, but it does not currently admit a 
polynomial-cost evaluation of CC wave functions. 

In this study, we introduce amplitude determinant coupled cluster theory (ADCC), an approach that seeks to maintain the properties of traditional CC, such as
size-consistency, while achieving polynomial cost and variational evaluation through VMC.
The central idea is to redefine the cluster expansion so that each configuration in the expansion has a coefficient given by a determinant of cluster amplitudes.
The thinking is that a determinant has the rare ability to map a combinatorially large sum of terms (the likes of which arise when one wants to consider all
possible excitation pathways into a given configuration) into an object that can be evaluated for a polynomially scaling cost.
While many choices for the matrix whose determinant is to be employed can be imagined, we seek in this study to develop what is likely the simplest example in which
we follow the pairwise doubles approach of pCCD, whose cluster expansion suggests a particularly simple choice for the matrix in question.
The resulting ADCC with pairwise doubles (pADCCD) proves to be remarkably similar to its non-variational cousin, motivating future investigations into
more general ADCC expansions.

This paper is structured as follows. We begin with a review of the pCCD method and explain the reason why it is incompatible with VMC. 
We then introduce pADCCD theory, its difference from pCCD, and prove that pADCCD is rigorously size-consistent. We will also explain the use of linear 
method\cite{Umrigar:2005:lm} technology to optimize its parameters. 
Having laid out the general formalism, we end our theory section by introducing orbital optimization of pADCCD with 
reduced density matrices (RDMs) and the Newton-Raphson algorithm, after which we conclude our theoretical analysis by discussing the scaling of the method. Results are 
presented for the bond stretching potentials of LiH, HF, H$_2$O and N$_2$, along with the comparison between canonical and our optimized orbitals, and a
size consistency check. We conclude with a summary of our findings and comments on the future development of pADCCD.

\section{Theory}
\label{sec:theory}

\subsection{CC Incompatibility with VMC}
\label{sec::cc_vmc}

In variational Monte Carlo, the energy expression is
\begin{align}
\label{eqn:vmc_energy}
E=\frac { \left< { \Psi  }|{ H }|{ \Psi  } \right>  }{ \left< { \Psi  }|\Psi  \right>  } =\sum _{ \vec{ n }  }^{  }{ \frac { { |\left< { \vec { n }  }|{ \Psi  } \right> | }^{ 2 } }{ \left< { \Psi  }|{ \Psi  } \right>  } \frac { \left< { \vec { n }  }|{ H }|{ \Psi  } \right>  }{ \left< { \vec { n }  }|{ \Psi  } \right>  }  } 
\end{align}
which may be approximated by sampling a set of occupations $\xi$ from the wave function's probability distribution 
${ |\left< { \vec { n } }|{ \Psi  } \right> | }^{ 2 }/\left< { \Psi  }|{ \Psi  } \right> $. The energy is then estimated as an average of local energies. 
\begin{align}
\label{eqn:mean_le}
E=\frac { 1 }{ { n }_{ s } } \sum _{ \vec { n } \in \xi  }^{  }{ \frac { \left< { \vec { n }  }|{ H }|{ \Psi  } \right>  }{ \left< { \vec { n }  }|{ \Psi  } \right>  }  } =\frac { 1 }{ { n }_{ s } } \sum _{ \vec { n } \in \xi  }^{  }{ { E }_{ L }\left( \vec { n }  \right)  } 
\end{align}
with $n_s$ being the number of Monte Carlo samples in $\xi$.

For our purposes, it is convenient to write the $ab$ $initio$ Hamiltonian as
\begin{align}
\label{eqn:hamiltonian}
\begin{split}
&H=\sum _{ pq }^{  }{ { t }_{ pq }\left( { a }_{ p }^{ \dagger  }{ a }_{ q }+{ a }_{ \overline { p }  }^{ \dagger  }{ a }_{ \overline { q }  } \right)  } \\
&+\sum _{ pqrs }^{  }{ \left( { pq }|{ rs } \right) \left( \frac { 1 }{ 2 } { a }_{ p }^{ \dagger  }{ a }_{ q }{ a }_{ r }^{ \dagger  }{ a }_{ s }+\frac { 1 }{ 2 } { a }_{ \overline { p }  }^{ \dagger  }{ a }_{ \overline { q }  }{ a }_{ \overline { r }  }^{ \dagger  }{ a }_{ \overline { s }  }+{ a }_{ p }^{ \dagger  }{ a }_{ q }{ a }_{ \overline { r }  }^{ \dagger  }{ a }_{ \overline { s }  } \right)  } 
\end{split}
\end{align}
where $\left( pq|{ rs } \right) $ are the usual two-electron coulomb integrals in $\left( 11|{ 22 } \right) $ order, ${ a }_{ p }^{ \dagger  }$ 
(${ a }_{ \overline { p }  }^{ \dagger  }$) and ${ a }_{ p }$(${ a }_{ \overline { p }  }$) are creation and destruction operators of an $\alpha$($\beta$) electron in the 
$p$th spatial orbital, and $t_{pq}$ are modified one-electron integrals,
\begin{align}
\label{eqn:moei}
{ t }_{ pq }={ h }_{ pq }-\frac { 1 }{ 2 } \sum _{ r }^{  }{ \left( { pr }|{ rq } \right)  } 
\end{align}
where $h_{pq}$ are the standard one-electron integrals. 

From the energy and Hamiltonian expressions, we see that if an ansatz can evaluate $\left< { \vec { n }  }|{ \Psi  } \right> $ efficiently, this is sufficient for use 
with VMC in Fock space. 
For the pCCD ansatz with RHF as reference, its cluster amplitudes can be write in matrix form
\begin{align}
\label{eqn:pccd_cluster}
T=\begin{pmatrix} { t }_{ i\overline { i }  }^{ a\overline { a }  } & { t }_{ j\overline { j }  }^{ a\overline { a }  } & { t }_{ k\overline { k }  }^{ a\overline { a }  } & ... \\ { t }_{ i\overline { i }  }^{ b\overline { b }  } & { t }_{ j\overline { j }  }^{ b\overline { b }  } & { t }_{ k\overline { k }  }^{ b\overline { b }  } & ... \\ { t }_{ i\overline { i }  }^{ c\overline { c }  } & { t }_{ j\overline { j }  }^{ c\overline { c }  } & { t }_{ k\overline { k }  }^{ c\overline { c }  } & ... \\ ... & ... & ... & ... \end{pmatrix}
\end{align}
in which index $i, j, k$ represent occupied orbitals in the reference and $a, b, c$ represent virtual orbitals in the reference. 

The amplitude expression for pCCD ansatz can be written as
\begin{align}
\left< { \vec { n }  }|{ { \Psi  }_{ pCCD } } \right>
& = \left< { \vec { n }  }|{ \mathrm{exp}\left( \hat { T }  \right)  }|{ RHF } \right> \notag \\
& =   \left< { \vec { n }  }|{ RHF } \right>
    + \left< { \vec { n }  }|{ \hat { T }  }|{ RHF } \right> \notag \\
& \quad  + \frac { 1 }{ 2 } \left< { \vec { n }  }|{ { \hat { T }  }^{ 2 } }|{ RHF } \right> + ...
\label{eqn:pccd_amp}
\end{align}
in which $\hat { T }$ is the cluster operator
\begin{align}
\label{eqn:pccd_cluster_ope}
\hat { T }=\sum _{ ia }^{  }{ { t }_{ i\overline { i }  }^{ a\overline { a }  }{ a }_{ a }^{ \dagger  }{ a }_{ i }{ a }_{ \overline { a }  }^{ \dagger  }{ a }_{ \overline { i }  } } 
\end{align}

Suppose the occupation number vector is a quadruply excited configuration, ${ \vec { n }  }_{ ij }^{ ab }$, relative to the reference determinant. 
Then only the second order cluster operator term will not vanish. 
Therefore, 
\begin{align}
\left< { { \vec { n } }_{ ij }^{ ab } }|{ { \Psi  }_{ pCCD } } \right>
&= \frac { 1 }{ 2 } \times 2 \times \left( { t }_{ i\overline { i }  }^{ a\overline { a }  }{ t }_{ j\overline { j }  }^{ b\overline { b }  }+{ t }_{ j\overline { j }  }^{ a\overline { a }  }{ t }_{ i\overline { i }  }^{ b\overline { b }  } \right) \notag \\
&= \frac { 1 }{ 2 } \times 2 \times \mathrm{Perm}\begin{pmatrix} { t }_{ i\overline { i }  }^{ a\overline { a }  } & { t }_{ j\overline { j }  }^{ a\overline { a }  } \\ { t }_{ i\overline { i }  }^{ b\overline { b }  } & { t }_{ j\overline { j }  }^{ b\overline { b }  } \end{pmatrix} \notag \\
&=\mathrm{Perm}\left[ T\left( { \vec { n } }_{ ij }^{ ab } \right)  \right] 
\label{eqn:pccd_amp_ex}
\end{align}
in which ``$\mathrm{Perm}$'' represents the permanent of a matrix and the factor 2 comes from the number of ways of permuting $i$, $j$, $a$, $b$ indices, which cancels 
the ${ 1 }/{ 2 }$ from the exponentials Taylor series. 
This result generalizes, with the amplitude of pCCD for a given occupation number 
vector given by a permanent of an ${n}_{ex}$/2 by ${n}_{ex}$/2 part of the cluster amplitude matrix, with ${ n }_{ ex }$ the excitation level in the 
occupation number vector\cite{Ayers:2013:ap1rog}. For $N$th level excitations, there are $N$! ways to permute the indices, 
therefore the exponential's 1$/N$! is canceled and no constants appear before the permanent.  

Thus, the difficulty in evaluating a pCCD occupation number coefficient is equivalent to the evaluation of a permanent. However, there are no known  
polynomial cost methods for permanent evaluation, and so pCCD, like CC in general, is incompatible with VMC. 

\subsection{Determinant Amplitude}
\label{sec:det_amp}
We propose a different, but related ansatz, in which coefficients are defined to be determinants rather than permanents. Using the $n_{ex}$/2 by $n_{ex}$/2 matrix
\begin{align}
\label{eqn:antisym_t}
\tilde { T } \equiv \begin{pmatrix} \quad { t }_{ i\overline { i }  }^{ a\overline { a }  } & \quad { t }_{ j\overline { j }  }^{ a\overline { a }  } & { \quad t }_{ k\overline { k }  }^{ a\overline { a }  } & ... \\ -{ t }_{ i\overline { i }  }^{ b\overline { b }  } & { \quad t }_{ j\overline { j }  }^{ b\overline { b }  } & \quad { t }_{ k\overline { k }  }^{ b\overline { b }  } & ... \\ -{ t }_{ i\overline { i }  }^{ c\overline { c }  } & -{ t }_{ j\overline { j }  }^{ c\overline { c }  } & \quad { t }_{ k\overline { k }  }^{ c\overline { c }  } & ... \\ ... & ... & ... & ... \end{pmatrix}
\end{align}

we define the wave function coefficient to be
\begin{align}
\label{eqn:det_amp}
\left< { \vec { n } }|{ \Psi  } \right> =\mathrm{det}\left[ \tilde { T } \left( \vec { n } \right)  \right].
\end{align}
These determinants can be evaluated through LU decomposition with $\mathcal{O}\left( { n }_{ ex }^{ 3 } \right) $ cost, making this pADCCD ansatz compatible with VMC.   

We should note that our choice of determinant instead of permanent forms a different approximation than pCCD, and since pCCD is not exact, using a determinant is not 
necessarily better or worse. Moreover, as we will prove later, size consistency still holds, and so one of CC's most important properties is maintained. 
And since we make the lower triangle negative, the coefficients obtained from pADCCD match those of pCCD through quadruple excitations. Unlike pCCD, pADCCD cannot be 
written in the compact $\mathrm{exp}\left( \hat { T }  \right) $ form, but with size consistency nonetheless maintained, this seems a minor inconvenience. 

\subsection{Energy Expression}
\label{sec:energy}
We will derive the energy expression for pADCCD in this subsection. Before considering our particular wave function, notice that the subset of 
Hamiltonian terms that contain only number and hole operators (i.e. ${ a }_{ i }^{ \dagger  }{ a }_{ i }$ and ${ a }_{ b }^{ \dagger  }{ a }_{ b }$)
will add a wave-function-independent contribution to the local energy,
\begin{align}
\label{eqn:wfn_ind_en}
{ E }_{ 0 }\left( \vec { n } \right)
&=
    \sum _{ i }^{  }{ { t }_{ ii } }
  + \sum _{ \overline { i }  }^{  }{ { t }_{ \overline { i } \overline { i }  } }
  + \sum _{ i\overline { j }  }^{  }{ \left( { ii }|{ \overline { j } \overline { j }  } \right)  } \notag \\
& \quad
  + \frac { 1 }{ 2 }
  \Bigg[
     \sum _{ ij }^{  }{ \left( { ii }|{ jj } \right)  }
    +\sum _{ ia }^{  }{ \left( { ia }|{ ai } \right)  } \notag \\
    & \qquad \qquad
    +\sum _{ \overline { i } \overline { j }  }^{  }{ \left( { \overline { i } \overline { i }  }|{ \overline { j } \overline { j }  } \right)  }
    +\sum _{ \overline { i } \overline { a }  }^{  }{ \left( { \overline { i } \overline { a }  }|{ \overline { a } \overline { i }  } \right)  }
  \Bigg] 
\end{align}

The remaining one-electron terms in the Hamiltonian of the type ${ a }_{ i }^{ \dagger  }{ a }_{ a }$ will not contribute to the local energy as it breaks seniority symmetry and pADCCD is strictly seniority zero. 
Likewise, the only two-electron terms in the Hamiltonian that will give a non-vanishing contribution to the local energy are the terms of the 
type ${ a }_{ i }^{ \dagger  }{ a }_{ a }{ a }_{ \overline { i }  }^{ \dagger  }{ a }_{ \overline { a }  }$. Therefore, the expression for the local energy is,
\begin{align}
E\left( \vec { n } \right)
& ={ E }_{ 0 }\left( \vec { n } \right) +\sum _{ ia }^{  }{ \left( { ia }|{ \overline { i } \overline { a }  } \right) \frac { \left< { \vec { n } }|{ { a }_{ i }^{ \dagger  }{ a }_{ a }{ a }_{ \overline { i }  }^{ \dagger  }{ a }_{ \overline { a }  } }|{ \Psi  } \right>  }{ \left< \vec{ n }|{ \Psi  } \right>  }  } \notag \\
& ={ E }_{ 0 }\left( \vec { n }  \right) +\sum _{ ia }^{  }{ \left( { ia }|{ \overline { i } \overline { a }  } \right) \frac { \left< { { \vec { n }  }_{ i\overline { i }  }^{ a\overline { a }  } }|{ \Psi  } \right>  }{ \left< \vec{ n }|{ \Psi  } \right>  }  } \notag \\
& ={ E }_{ 0 }\left( \vec { n }  \right) +\sum _{ ia }^{  }{ \left( { ia }|{ \overline { i } \overline { a }  } \right) \frac { \mathrm{det}\left[ \tilde { T } \left( { \vec { n }  }_{ i\overline { i }  }^{ a\overline { a }  } \right)  \right]  }{ \mathrm{det}\left[ \tilde { T } \left( \vec { n }  \right)  \right]  }  } 
\label{eqn:le}
\end{align}
in which $\left| { \vec { n } }_{ i\overline { i } }^{ a\overline { a } }  \right> $ is obtained by exciting two electrons ($\alpha$ and $\beta$)
from the $ith$ occupied orbital to the $ath$ virtual orbital in $\left |\vec{n} \right> $.

\subsection{Proof of Size-Consistency}
\label{sec:proof_sc}
In this subsection, we will prove that our pADCCD ansatz is strictly size consistent. The definition of size-consistency states that the energy calculated with 
two non-interacting systems A and B together as a ``super system'' should be equal to the sum of the energies of systems A and B calculated separately. To prove this, 
consider two non-interacting systems A and B. Then the coefficient matrix becomes block diagonal
\begin{align}
\label{eqn:block_diagonal}
\tilde { T } =\begin{pmatrix} { \tilde { T }  }_{ AA } & 0 \\ 0 & { \tilde { T }  }_{ BB } \end{pmatrix}
\end{align}
since there are no ``cross-excitations'' between the two systems. Then due to the properties of determinants, we have
\begin{align}
\label{eqn:productive_separable}
\left< { { \vec { n } }_{ AB } }|{ { \Psi  }_{ AB } } \right> =\left< { { \vec { n } }_{ A } }|{ { \Psi  }_{ A } } \right> \left< { { \vec { n } }_{ B } }|{ { \Psi  }_{ B } } \right> 
\end{align}
and it follows that,
\begin{align}
\label{eqn:ab_innerproduct}
\begin{split}
&\left< { { \Psi  }_{ AB } }|{ { \Psi  }_{ AB } } \right> =\sum _{ { \vec { n }  }_{ AB } }^{  }{ \left< { { \Psi  }_{ AB } }|{ { \vec { n }  }_{ AB } } \right> \left< { { \vec { n }  }_{ AB } }|{ { \Psi  }_{ AB } } \right>  } \\
&=\sum _{ { \vec { n }  }_{ AB } }^{  }{ { \left| \left< { { \Psi  }_{ A } }|{ { \vec { n }  }_{ A } } \right>  \right|  }^{ 2 }{ \left| \left< { { \Psi  }_{ B } }|{ { \vec { n }  }_{ B } } \right>  \right|  }^{ 2 } } \\
&=\left< { { \Psi  }_{ A } }|{ { \Psi  }_{ A } } \right> \left< { { \Psi  }_{ B } }|{ { \Psi  }_{ B } } \right> \\
\end{split}
\end{align}
Thus the energy expression becomes
\begin{align}
&{ E }_{ AB }=\sum _{ { \vec { n }  }_{ AB } }^{  }{ \frac { { \left| \left< { { \Psi  }_{ AB } }|{ { \vec { n }  }_{ AB } } \right>  \right|  }^{ 2 } }{ \left< { { \Psi  }_{ AB } }|{ { \Psi  }_{ AB } } \right>  } \frac { \left< { { \vec { n }  }_{ AB } }|{ { H }_{ A }+{ H }_{ B } }|{ { \Psi  }_{ AB } } \right>  }{ \left< { { \vec { n }  }_{ AB } }|{ \Psi  }_{ AB } \right>  }  } \notag \\
&=\sum _{ { \vec { n }  }_{ AB } }^{  }
\frac { { \left| \left< { { \Psi  }_{ AB } }|{ { \vec { n }  }_{ AB } } \right>  \right|  }^{ 2 } }{ \left< { { \Psi  }_{ AB } }|{ { \Psi  }_{ AB } } \right>  }
\Bigg( \frac { \left< { { \vec { n }  }_{ A } }|{ { H }_{ A } }|{ { \Psi  }_{ A } } \right>  }{ \left< { { \vec { n }  }_{ A } }|{ { \Psi  }_{ A } } \right>  } \frac { \left< { { \vec { n }  }_{ B } }|{ { \Psi  }_{ B } } \right>  }{ \left< { { \vec { n }  }_{ B } }|{ { \Psi  }_{ B } } \right>  } \notag \\
& \qquad \qquad \qquad \qquad \qquad \quad
      +\frac { \left< { { \vec { n }  }_{ B } }|{ { H }_{ B } }|{ { \Psi  }_{ B } } \right>  }{ \left< { { \vec { n }  }_{ B } }|{ { \Psi  }_{ B } } \right>  } \frac { \left< { { \vec { n }  }_{ A } }|{ { \Psi  }_{ A } } \right>  }{ \left< { { \vec { n }  }_{ A } }|{ { \Psi  }_{ A } } \right>  }
\Bigg)
\notag \\
&=\sum _{ { \vec { n }  }_{ AB } }^{  }{ \frac { { \left| \left< { { \Psi  }_{ A } }|{ { \vec { n }  }_{ A } } \right>  \right|  }^{ 2 }{ \left| \left< { { \Psi  }_{ B } }|{ { \vec { n }  }_{ B } } \right>  \right|  }^{ 2 } }{ \left< { { \Psi  }_{ A } }|{ { \Psi  }_{ A } } \right> \left< { { \Psi  }_{ B } }|{ { \Psi  }_{ B } } \right>  } \left( { E }_{ { L }_{ A } }\left( { \vec { n }  }_{ A } \right) +{ E }_{ { L }_{ B } }\left( { \vec { n }  }_{ B } \right)  \right)  } \notag \\
&=\sum _{ { \vec { n }  }_{ A } }^{  }{ \frac { { \left| \left< { { \Psi  }_{ A } }|{ { \vec { n }  }_{ A } } \right>  \right|  }^{ 2 } }{ \left< { { \Psi  }_{ A } }|{ { \Psi  }_{ A } } \right>  } { E }_{ { L }_{ A } }\left( { \vec { n }  }_{ A } \right)  } +\sum _{ { \vec { n }  }_{ B } }^{  }{ \frac { { \left| \left< { { \Psi  }_{ B } }|{ { \vec { n }  }_{ B } } \right>  \right|  }^{ 2 } }{ \left< { { \Psi  }_{ B } }|{ { \Psi  }_{ B } } \right>  } { E }_{ { L }_{ B } }\left( { \vec { n }  }_{ B } \right)  } \notag \\
&={ E }_{ A }+{ E }_{ B }
\label{eqn:super_energy}
\end{align}
and so pADCCD is size consistent. 

\subsection{Derivative Ratios}
Before discussing the variational minimization of the energy, we first lay the groundwork by developing efficient evaluations of terms that we call derivative ratios, which 
will make the optimization relatively simple to describe. Defining derivative 
notation $\left| { \Psi  }^{ x } \right> \equiv \partial \left| \Psi  \right> /\partial { \mu  }_{ x }$, where ${ \mu  }_{ x }$ is the xth wave function parameter, we first 
consider the ``bare'' derivative ratio
\begin{align}
\label{eqn:der_rat}
\mathcal{ D }_{ \vec { n } }\left( { \mu  }_{ x } \right) \equiv \frac { \left< { \vec { n } }|{ { \Psi  }^{ x } } \right>  }{ \left< { \vec { n } }|{ { \Psi  } } \right>  } 
\end{align}

For amplitudes, these ratios are:
\begin{align}
\label{eqn:amp_rat}
\mathcal{ D }_{ \vec { n } }\left( { t }_{ i\overline { i }  }^{ a\overline { a }  } \right) \equiv \frac { \mathrm{det}\left[ \tilde { T } \left( \vec { n } \right)  \right] Tr\left[ \Theta \left( \vec { n } \right) \frac { \partial \tilde { T } \left( \vec { n } \right)  }{ \partial { t }_{ i\overline { i }  }^{ a\overline { a }  } }  \right]  }{ \mathrm{det}\left[ \tilde { T } \left( \vec { n } \right)  \right]  } ={ \Theta \left( \vec { n } \right)  }_{ a\overline { a }  }^{ i\overline { i }  }
\end{align}
in which $\Theta \left( \vec { n } \right) $ is the inverse of $\tilde { T } \left( \vec { n } \right) $. 

In addition to the bare derivative ratios, we also define the energy derivative ratios,
\begin{align}
\label{eqn:en_der}
\mathcal{ G }_{ \vec { n } }\left( { \mu  }_{ x } \right) \equiv \frac { \left< { \vec { n } }|{ H }|{ { \Psi  }^{ x } } \right>  }{ \left< { \vec { n } }|{ \Psi  } \right>  } 
\end{align}
which are related to derivatives of the local energy by
\begin{align}
\label{eqn:en_der2}
\mathcal{ G }_{ \vec { n } }\left( { \mu  }_{ x } \right) =\frac { \partial { E }_{ L }\left( \vec { n } \right)  }{ \partial { \mu  }_{ x } } +\mathcal{ D }_{ \vec { n } }\left( { \mu  }_{ x } \right) { E }_{ L }\left( \vec { n } \right) 
\end{align}
Based on the expression of local energy, its derivative is:
\begin{align}
\label{eqn:le_der}
\frac { \partial { E }_{ L }\left( \vec { n } \right)  }{ \partial { t }_{ j\overline { j }  }^{ b\overline { b }  } }=\sum _{ ia }^{  }{ \left( { ia }|{ \overline { i } \overline { a }  } \right) \frac { \mathrm{det}\left[ \tilde { T } \left( { \vec { n }  }_{ i\overline { i }  }^{ a\overline { a }  } \right)  \right]  }{ \mathrm{det}\left[ \tilde { T } \left( \vec { n }  \right)  \right]  } \left( { \Theta \left( \vec { n }  \right)  }_{ b\overline { b }  }^{ j\overline { j }  }-{ \Theta \left( { \vec { n }  }_{ i\overline { i }  }^{ a\overline { a }  } \right)  }_{ b\overline { b }  }^{ j\overline { j }  } \right)  } 
\end{align}
and the evaluation of local energy derivatives are similar to that of bare derivative ratios.

\subsection{Amplitude Optimization by Linear Method}
We variationally optimize the energy of the pADCCD wave function using the linear method (LM)\cite{Nightingale:2001:linear_method, UmrTouFilSorHen-PRL-07}. The LM works by repeatedly solving 
the Schr\"{o}dinger equation in a special subspace 
of the full Hilbert space defined by the span of the wave function and its first derivatives. More precisely, we construct the generalized eigenvalue problem
\begin{align}
\label{eqn:eigen}
\sum _{ y\in \left\{ 0,1,... \right\}  }^{  }{ \left< { { \Psi  }^{ x } }|{ H }|{ { \Psi  }^{ y } } \right> { c }_{ y } } =\lambda \sum _{ y\in \left\{ 0,1,... \right\}  }^{  }{ \left< { { \Psi  }^{ x } }|{ { \Psi  }^{ y } } \right> { c }_{ y } } 
\end{align}
where $\left |{ \Psi  }^{ x } \right> $ and $\left| { \Psi  }^{ y } \right> $ are shorthand for derivatives of $\left| { \Psi  } \right> $ with respect to the xth and yth
wave function parameters ${ \mu  }_{ x }$ and ${ \mu  }_{ y }$, respectively, and $\left |{ \Psi  }^{ 0 } \right> \equiv \left |\Psi  \right> $. After solving this eigenvalue
problem for $c$, one updates the parameters by 
\begin{align}
\label{eqn:param_update}
{ \mu  }_{ x }\leftarrow { \mu  }_{ x }+{ c }_{ x }/{ c }_{ 0 }
\end{align}

The Hamiltonian and overlap matrices are built by Monte Carlo sampling
\begin{align}
\label{eqn:matrix_build}
\begin{split}
&\sum _{ \vec { n }\in \xi  }^{  }{ \sum _{ y\in \left\{ 0,1,... \right\}  }^{  }{ \frac { \left< { { \Psi  }^{ x } }|{ \vec { n } } \right>  }{ \left< { \Psi  }|{ \vec { n } } \right>  } \frac { \left< { \vec { n } }|{ H }|{ { \Psi  }^{ y } } \right>  }{ \left< { \vec { n } }|{ \Psi  } \right>  } { c }_{ y } }  } \\
&=\lambda \sum _{ \vec { n }\in \xi  }^{  }{ \sum _{ y\in \left\{ 0,1,... \right\}  }^{  }{ \frac { \left< { { \Psi  }^{ x } }|{ \vec { n } } \right>  }{ \left< { \Psi  }|{ \vec { n } } \right>  } \frac { \left< { \vec { n } }|{ { \Psi  }^{ y } } \right>  }{ \left< { \vec { n } }|{ \Psi  } \right>  } { c }_{ y } }  }
\end{split}
\end{align}
Inspecting the above Monte Carlo approximations to the Hamiltonian and overlap matrices in the first derivative subspace makes clear that only derivative ratios 
$\mathcal{ D }_{ n }\left( { \mu  }_{ x } \right) $ and $\mathcal{ G }_{ n }\left( { \mu  }_{ x } \right) $ from Eqs. \ref{eqn:der_rat} and \ref{eqn:en_der} are needed. 
Therefore, the LM can be applied efficiently to optimize the amplitudes by evaluating these ratios. 

\subsection{Orbital Optimization}
Like DOCI and pCCD, pADCCD is not invariant to the choice of orbital basis, since all open-shell determinants are neglected. Thus optimal orbitals need to 
be found to fully minimize the energy. In this section we introduce an orbital optimization method for pADCCD. First consider the one-body anti-Hermitian operator
\begin{align}
\label{eqn:orb_rot_op}
\sum _{ p>q }^{  }{ \sum _{ \sigma  }^{  }{ { \kappa  }_{ pq }\left( { a }_{ p\sigma  }^{ \dagger  }{ a }_{ q\sigma  }-{ a }_{ q\sigma  }^{ \dagger  }{ a }_{ p\sigma  } \right)  }  } 
\end{align}
which, when exponentiated, creates unitary orbital rotations (here $\sigma $ indexes spin). 

Given this rotation operator, we can generalize the energy to be
\begin{align}
\label{eqn:en_orb}
E\left( \kappa  \right) =\left< { \Psi  }|{ e }^{ -\kappa  }H{ e }^{ \kappa  }|{ \Psi  } \right> 
\end{align}

We expand this energy to second order in $\vec{\kappa}$,
\begin{align}
E\left( \vec{\kappa}  \right) \simeq E\left( 0 \right)
&+{ \vec{\kappa}  }^{ T }
{ \left( \frac { \partial E\left( \vec{\kappa}  \right)  }{ \partial \vec{\kappa}  }  \right)  }_{ \vec{\kappa} =0 } \notag \\
& \quad +\frac { 1 }{ 2 } { \vec{\kappa}  }^{ T }{ \left( \frac { { \partial  }^{ 2 }E\left( \vec{\kappa}  \right)  }{ \partial { \vec{\kappa}  }^{ 2 } }  \right)  }_{ \vec{\kappa} =0 }\vec{\kappa} 
\label{eqn:2nd_e}
\end{align}
where we work at $\vec{\kappa} =0$ by transforming the basis in which we express the Hamiltonian (i.e. by transforming the one- and two-electron integrals).  

We define the energy gradient and hessian,
\begin{align}
\label{eqn:en_gra}
{ \omega  }_{ pq }=\frac { \partial E\left( \kappa  \right)  }{ \partial { \kappa  }_{ pq } } \\
{ A }_{ pq,rs }=\frac { { \partial  }^{ 2 }E\left( \kappa  \right)  }{ \partial { \kappa  }_{ pq }\partial { \kappa  }_{ rs } } 
\end{align}

both of which are functions of the one- and two-electron reduced density matrices
\begin{align}
\label{eqn:rdm}
\begin{split}
&{ \gamma  }_{ pq }\equiv \left< { \Psi  }|{ { { a }_{ p }^{ \dagger  }{ a }_{ q }+{ a }_{ \overline { p }  }^{ \dagger  }{ a }_{ \overline { q }  } } }|{ \Psi  } \right> \\
&{ \Gamma  }_{ pq }^{ rs }\equiv \left< { \Psi  }|\frac { 1 }{ 2 } { { { a }_{ p }^{ \dagger  }{ a }_{ r }{ a }_{ q }^{ \dagger  }{ a }_{ s }+\frac { 1 }{ 2 } { a }_{ \overline { p }  }^{ \dagger  }{ a }_{ \overline { r }  }{ a }_{ \overline { q }  }^{ \dagger  }{ a }_{ \overline { s }  }+{ a }_{ p }^{ \dagger  }{ a }_{ r }{ a }_{ \overline { q }  }^{ \dagger  }{ a }_{ \overline { s }  } } }|{ \Psi  } \right> 
\end{split}
\end{align}
We evaluate both $\bm{\gamma}$ and $\bm{\Gamma}$ the same way we evaluate the energy. 

It should be noted that the one-electron reduced density matrix $\gamma$ is diagonal in the basis in which we define the pairing. The two-electron reduced density 
matrix $\Gamma$ is also very sparse thanks to pADCCD being seniority zero. Detailed expressions for the orbital gradient and hessian can be found in the Appendix. 

Using $\vec{\omega}$ and $\bm{A}$, we can minimize the energy with respect to orbital rotations by Newton-Raphson,
\begin{align}
\label{eqn:nr}
\vec{\kappa} =-\bm{ A }^{ -1 }\vec{\omega} 
\end{align}
After each step, we update the MO-coefficients by 
\begin{align}
\label{eqn:basis_trans}
\bm{C}=\bm{ C }^{ \left( 0 \right)  }\bm{U}=\bm{ C }^{ \left( 0 \right)  }{ e }^{ \vec{\kappa}  }
\end{align}
where $\bm{ C }^{ \left( 0 \right)  }$ is the old MO-coefficients matrix, $\bm{C}$ is the new MO-coefficients matrix, and $\bm{U}$ is the MO-rotation matrix. We then recompute our 
MO integrals to work in the basis in which $\vec{\kappa}$ is again zero. 

In each iteration, we first perform a linear method update for the cluster amplitudes. 
Then we evaluate the one- and two-electron reduced density matrices with the updated wave function. Finally, $\vec{\omega}$ and $\bm{A}$ are built and NR is performed to 
update the orbital basis. 

\begin{table}[b]
\centering
\caption{Size consistency errors (absolute values) for well separated H$_2$ in kcal/mol.}
\label{tab:size_con_err}
\begin{tabular}{ c  r@{$\pm$}l  r@{.}l }
\hline\hline
\hspace{0mm} number of H$_2$ \hspace{0mm} & 
\multicolumn{2}{ c }{ \hspace{0mm} pADCCD \hspace{0mm} } &
\multicolumn{2}{ c }{ \hspace{0mm} CISD \hspace{0mm} }\\
\hline
 \hspace{0mm} 1 \hspace{0mm} & \hspace{0mm}  0.000&0.001 \hspace{0mm} &  0&00 \hspace{0mm} \\
 \hspace{0mm} 2 \hspace{0mm} & \hspace{0mm}  0.0001&0.0003 \hspace{0mm} &  0&64 \hspace{0mm} \\
 \hspace{0mm} 3 \hspace{0mm} & \hspace{0mm}  0.003&0.008 \hspace{0mm} &  1&21 \hspace{0mm} \\
 \hspace{0mm} 4 \hspace{0mm} & \hspace{0mm}  0.001&0.006 \hspace{0mm} & 1&73 \hspace{0mm} \\
 \hspace{0mm} 5 \hspace{0mm} & \hspace{0mm}  0.001&0.006 \hspace{0mm} & 2&21 \hspace{0mm} \\
\hline\hline
\end{tabular}
\end{table}

\subsection{Scaling}
Having presented our optimization method, we now analyze its cost. In each iteration, we need to loop over occupied and virtual orbitals to compute local energies,
RDMs, and derivative ratios. As the evaluation of the relevant determinants scale as ${ n }_{ ex }^{ 3 }$, in which $n_{ex}$ is the pair excitation level, 
the overall cost scales as ${ n }_{ s }{ n }_{ o }{ n }_{ u }{ n }_{ ex }^{ 3 }$, in which $n_s$, $n_o$ and $n_u$ are the number of samples, number of occupied and unoccupied orbitals in reference determinant, respectively.   
Although this $\mathcal{O}\left(N^6\right)$ scaling looks much steeper than the $\mathcal{O}\left(N^3\right)$ 
scaling of pCCD, one should note the highly excited configurations are rarely sampled, and so in many molecules the ${ n }_{ ex }^{ 3 }$ term in the scaling
may behave more like a constant. In such a regime, pADCCD's scaling may appear closer to $\mathcal{O}\left(N^3\right)$. 

At the end of each iteration, we need to reset $\hat { \kappa  } $ to 0 via a basis rotation. The one- and two-electron integrals needed to represent Hamiltonian in 
the new basis can be evaluated at an $\mathcal{O}\left( { N }^{ 5 } \right) $ cost. However, as the basis rotation is required only once per LM iteration, rather than 
once per sample, its cost is typically negligible compared to that of the sampling effort involved in optimizing cluster amplitudes.

\begin{figure}[t]
 \includegraphics[width=6.7cm, angle=270]{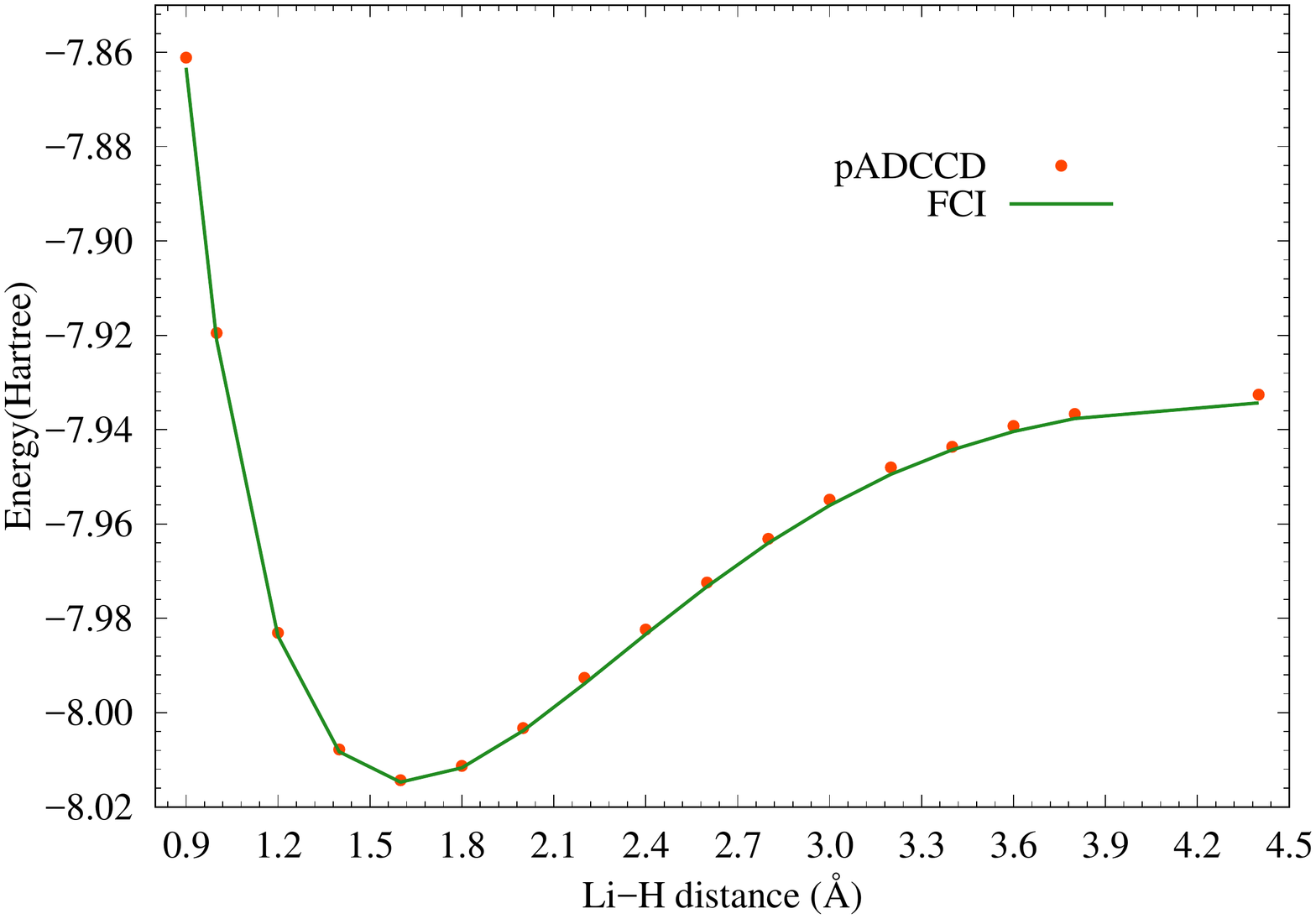}
 \caption{Dissociation of LiH in cc-pVDZ basis set}
 \label{fig:LiH}
\end{figure}

\section{Results}
\label{sec:results}

\subsection{Computational Details}
pADCCD results were obtained using our own software for VMC in Hilbert space, with one- and two-electron integrals for the Hamiltonian taken from PySCF\cite{Pyscf_brief}. 
The full configuration interaction (FCI) results were obtained from Molpro\cite{MOLPRO_brief} and CISD results from Psi4\cite{Psi4:2012}.
pCCD and DOCI results were kindly shared by Peter A. Limacher\cite{person_comm}. 
We froze N and O 1s orbitals at the RHF level. Sample size is taken to be 3.6$\times$10$^6$. All statistical uncertainties were converged to less than 0.01eV in all cases. 

\subsection{Size Consistency Check}

Before showing our examples, we first check the size consistency of pADCCD by calculating the energy of up to 5 well separated H$_2$ molecules in a STO-3G basis. 
As shown in Table \ref{tab:size_con_err}, the error per molecule does not grow with the increase of system size like in CISD. 
Instead, it remains zero within statistical uncertainty. This result demonstrates pADCCD's size consistency, as we proved in section \ref{sec:proof_sc}. 

\begin{figure}[t]
 \includegraphics[width=6.7cm, angle=270]{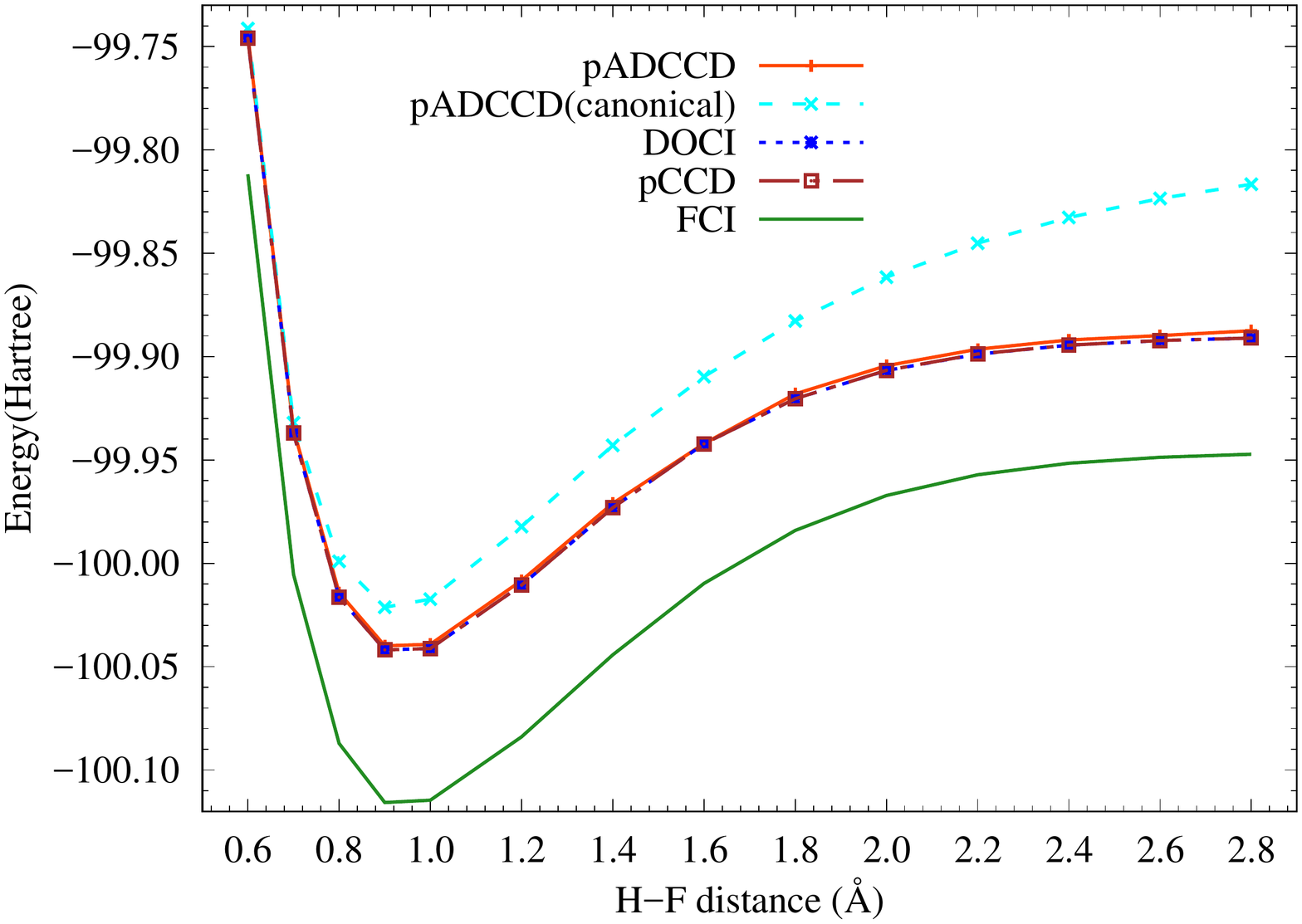}
 \caption{Dissociation of HF in 6-31G basis set}
 \label{fig:HF}
\end{figure}

\begin{figure}[b]
 \includegraphics[width=6.7cm, angle=270]{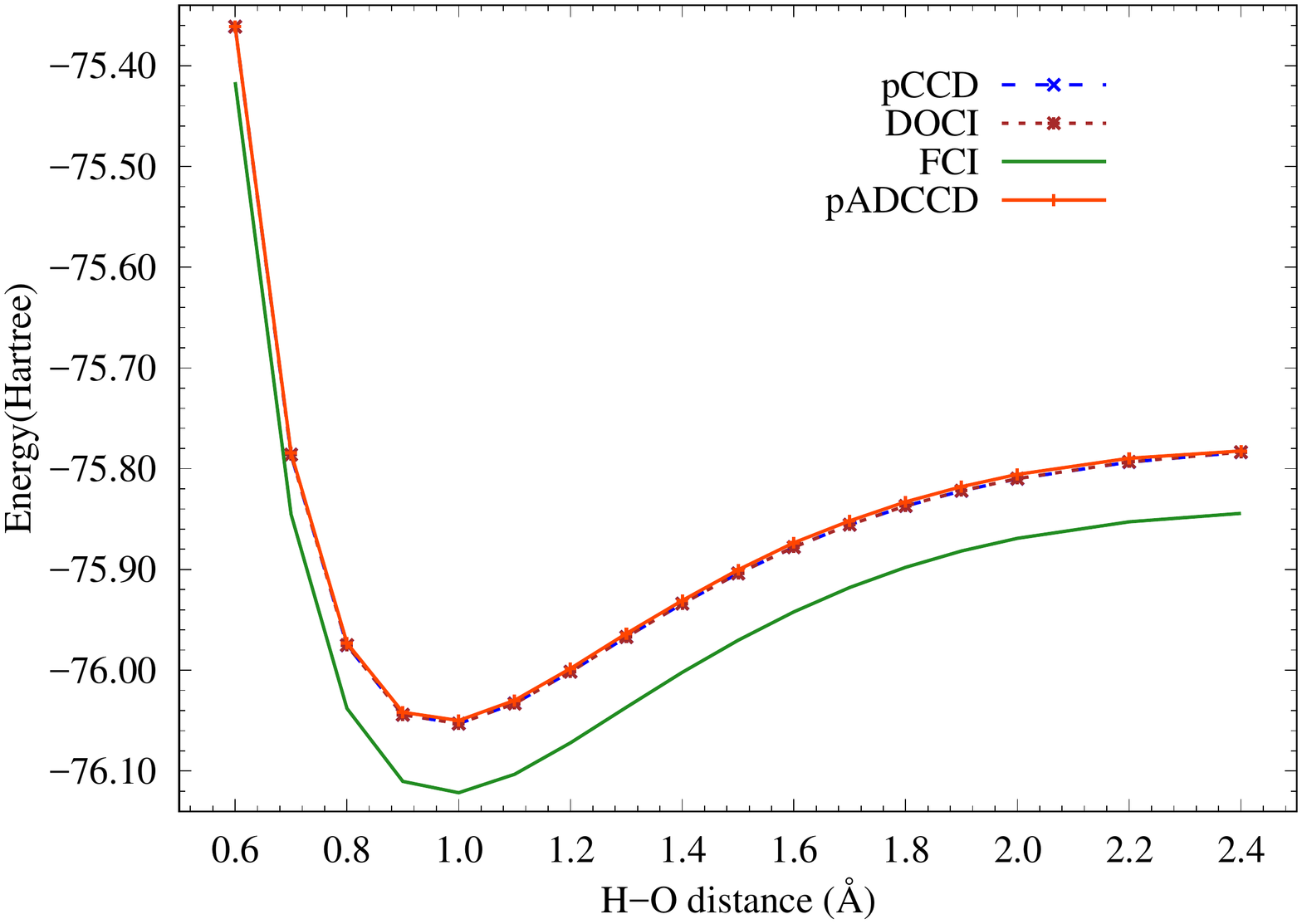}
 \caption{Symmetric Dissociation of H$_2$O in 6-31G basis set with a bond angle of 109.57$^{\circ}$}
 \label{fig:H2O}
\end{figure}

\subsection{LiH}

We begin our results with a simple example, the LiH molecule in the cc-pVDZ basis\cite{cc-pvdz}. The system has only two valence electrons, and both 
DOCI and pCCD deliver almost exact results compared to FCI. Due to our method's similarity with pCCD, we also expect nearly exact results. 
Figure \ref{fig:LiH} shows our results at 17 bond lengths between 0.9$\mathring { A } $ and 4.4$\mathring { A } $. The non-parallelarity error (NPE) defined as the 
difference between the largest and smallest error with respect to FCI along the potential surface is about 2 mE$_h$, confirming our expectations.

\subsection{HF}

\begin{figure*}
\centering
\subfloat{\includegraphics[width=1.6in]{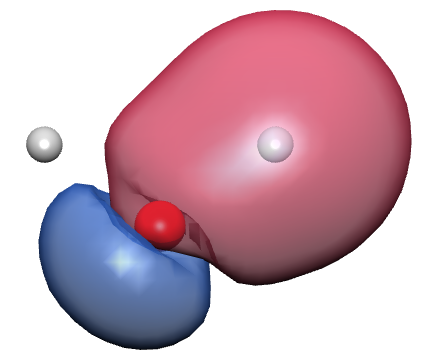}} 
\qquad
\subfloat{\includegraphics[width=1.6in]{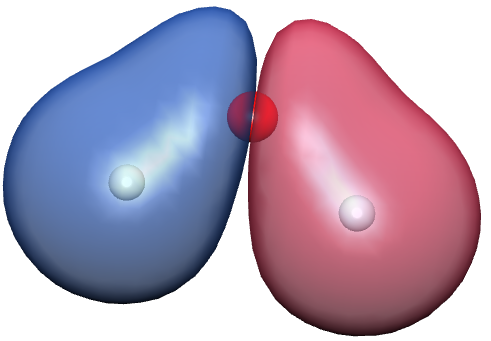}} 
\\
\subfloat{\includegraphics[width=1.6in]{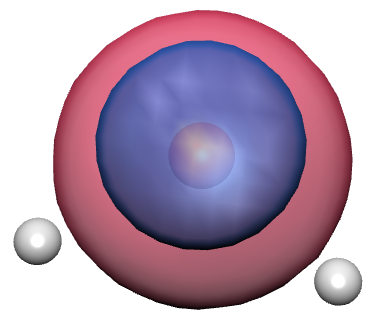}} 
\qquad
\subfloat{\includegraphics[width=1.6in]{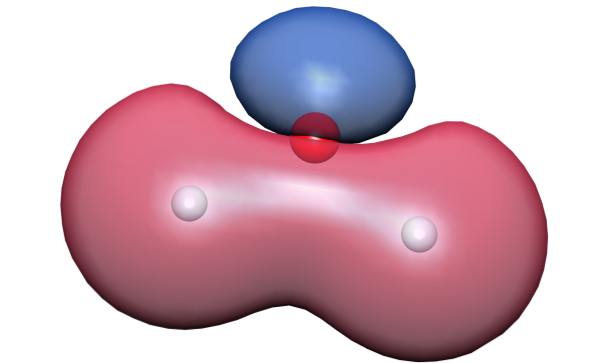}} 
\caption{Comparison of H$_2$O optimized and canonical orbitals.
         Clockwise from top left the orbitals are the optimized 4th, canonical 4th, canonical 5th, and optimized 5th.
        }
\label{fig:orb_com}
\end{figure*}

We next turn our attention to the dissociation of hydrogen flouride in a 6-31G basis\cite{6-31g}. Figure \ref{fig:HF} shows the absolute energy of pADCCD, 
along with DOCI, pCCD\cite{person_comm} and FCI results. The results show that pADCCD, DOCI and pCCD 
are energetically almost indistinguishable. A close analysis shows that the 
NPE of pADCCD and pCCD with respect to FCI is about 17 and 18 mE$_h$, respectively. Unlike LiH, where seniority zero based wave functions are near exact, we see a large
energy gap between all seniority zero based ansatzes and FCI. This is a remainder that while seniority zero wave functions are often effective for strong correlations, 
they do not capture all the details of weak correlation. 

In Figure \ref{fig:HF} we also plot the results using only RHF canonical orbitals and no further orbital optimization. 
As one can see, the results are quite poor when using RHF orbitals, especially when one stretches the bonds and the optimal orbitals become more and more localized. 

\subsection{H$_2$O}

Our next example is the symmetric double dissociation of H$_2$O, as shown in Figure \ref{fig:H2O}. Again, pADCCD provides nearly identical energies compared to 
DOCI and pCCD, and the NPE with respect to FCI is about 19 mE$_h$ for pADCCD and 16 mE$_h$ for pCCD. The coincidence of the pADCCD with pCCD and DOCI is
thus true not just for one pair of strongly correlated electrons, but for two pairs as well. However, we can still see from the plot that like pCCD and DOCI, 
a significant amount of dynamic correlation is clearly missing in all these methods. 

In order to show the importance of orbital optimization, we plot optimized and RHF canonical orbitals for H$_2$O at bond length 2.2$\mathring { A } $ 
in Figure \ref{fig:orb_com}. It is very clear that at this stretched geometry, the optimized orbitals are much more localized than canonical orbitals. This localization
is also seen in pCCD\cite{Scuseria:2011:seniority}. Indeed, the qualitative difference between optimized orbitals and canonical orbitals emphasizes the necessity of
orbital optimization. 

%

\begin{figure}[b]
 \includegraphics[width=6.7cm, angle=270]{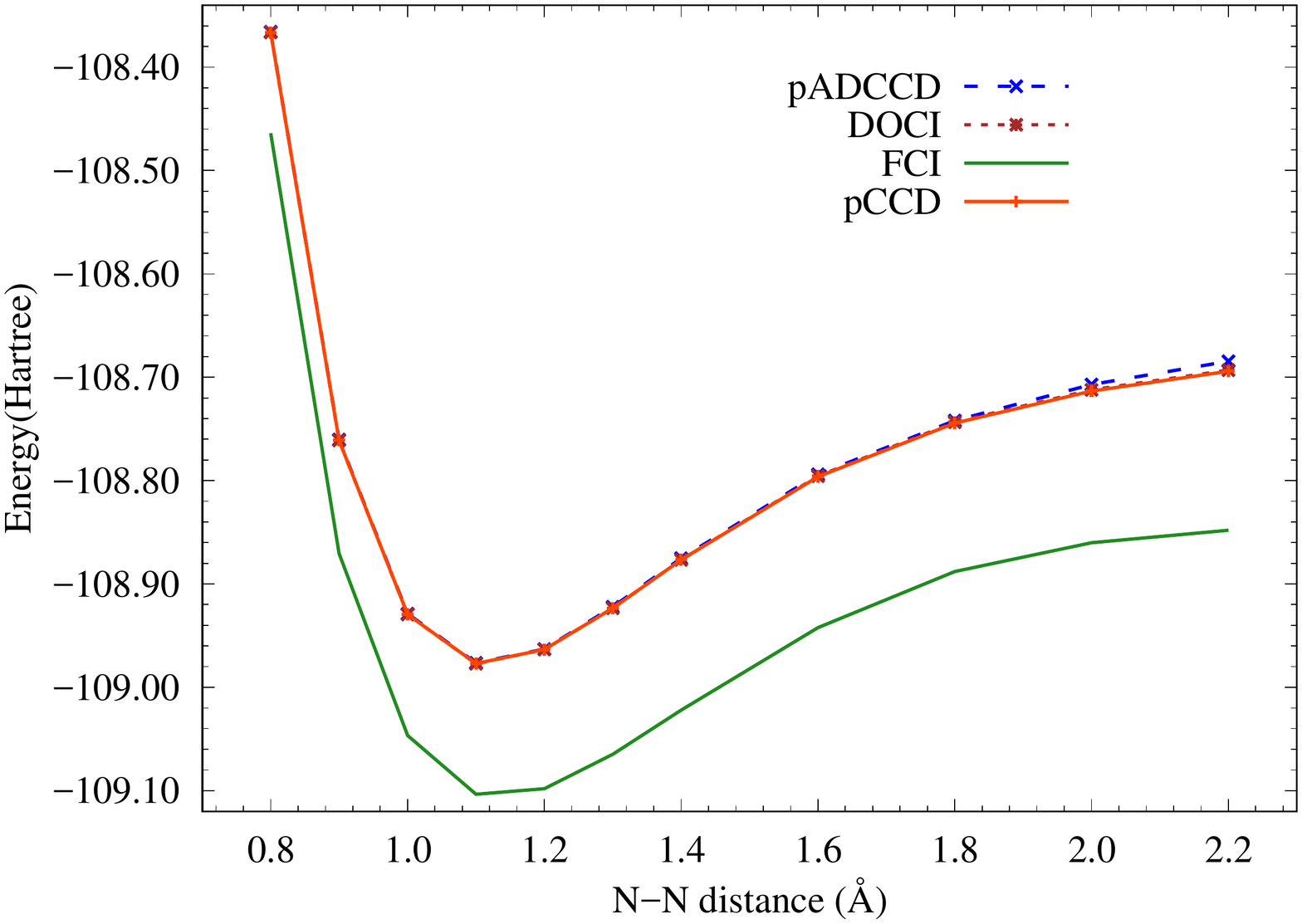}
 \caption{Dissociation of N$_2$ in 6-31G basis set}
 \label{fig:N2}
\end{figure}

\subsection{N$_2$}

Our final example is the dissociation of N$_2$.
As Figure \ref{fig:N2} reveals, the difference between pCCD and pADCCD is more noticeable in N$_2$, which is to be extected now that hextuples
(the first excitation level at which the ansatz forms are different) are needed for a qualitatively correct description of the dissociation.
Indeed, we find that disabling hextuple and higher excitations raises the pADCCD energy by 0.5eV.
Although their NPEs now differ noticeably, 56 mE$_h$ for pCCD versus 65 mE$_h$ for pADCCD, they are of the same order of magnitude and neither
are close to quantitative.
Achieving a more quantitative accuracy clearly requires a more flexible cluster expansion, but, as is well known \cite{}, this route leads
to qualitatively incorrect variational violations when pursued in a traditional CC approach.
As variational violations are not possible in a VMC-based approach, it will be interesting in future to investigate more flexible expansions
within the amplitude determinant framework.

\section{Conclusions}
\label{sec:conclusions}
We have presented amplitude determinant coupled cluster with pairwise doubles (pADCCD) as a variational cousin to pCCD.
Unlike the permanent-based coefficients of pCCD, pADCCD defines its expansion coefficients as amplitude determinants.
Combined with variational Monte Carlo methods, this choice produces a method that is exact for an electron pair, size-consistent,
polynomial cost, \textit{and} variational.
Initial tests on the dissociations of LiH, HF, H$_2$O and N$_2$ reveal that pADCCD and pCCD produce similar results,
suggesting that the leading approximation in both theories is their limitation to the seniority zero sector rather than the choice
of permanent versus determinant for defining the cluster expansion.

Like pCCD and other seniority zero approaches, pADCCD proves effective for describing some strong electron correlations but is unable to deliver
quantitative accuracy, a difficulty that may in future be addressed in two different ways.
First, one may seek to increase the flexibility of the cluster expansion.
We know that generalizing pCCD into CCSD greatly improves the recovery of weak correlation effects near equilibrium, but that it also leads to 
unacceptable variational violations as bonds are stretched.
Analogous generalizations of pADCCD cannot suffer this problem, and so exploring more sophisticated amplitude-determinant-based cluster expansions
is one attractive option.
Another approach would be to use an amplitude-determinant-based expansion as the trial function in projector Monte Carlo,
which the low per-sample cost of pADCCD for low-lying configurations suggests may be an especially effective pairing.
Of course, these avenues are not mutually exclusive, and we look forward to investigating both in future research.

\section{Acknowledgement}
The authors thank Peter A. Limacher for sharing with us his raw pCCD and DOCI data. We also acknowledge funding from the Office of Science, Office of Basic
Energy Sciences, the US Department of Energy, Contract No. DE-AC02-05CH11231. Calculations were performed using the Berkeley Research Computing Savio cluster. 

\vspace{5mm}


\noindent
\textbf{\large Appendix}

\vspace{2mm}

For completeness, we include here expression for the pADCCD orbital rotation gradient and hessian. These should provide 
everything needed for the Newton-Raphson algorithm we use for orbital optimization. The expressions are similar to those in the pCCD paper\cite{Scuseria:2014:pccd2}.

The energy can be written as
\begin{align}
\label{eqn:en_orb_rot_app}
E\left( \kappa  \right) =\left< { \Psi  }|{ { e }^{ -\kappa  }H{ e }^{ \kappa  } }|{ \Psi  } \right> 
\end{align}
with 
\begin{align}
\label{eqn:orb_rot_app}
\kappa =\sum _{ p>q }^{  }{ \sum _{ \sigma  }^{  }{ { \kappa  }_{ pq }\left( { a }_{ p\sigma  }^{ \dagger  }{ a }_{ q\sigma  }-{ a }_{ q\sigma  }^{ \dagger  }{ a }_{ p\sigma  } \right)  }  } 
\end{align}
where we evaluate the orbital rotation unitary transformation as exp$\left( \kappa  \right) $ and transform the integrals into this new basis. 

The orbital gradient is
\begin{align}
\label{eqn:orb_gra}
\begin{split}
&{ \left( \frac { \partial E\left( \kappa  \right)  }{ \partial { \kappa  }_{ pq } }  \right)  }_{ \kappa =0 }
=\mathcal{ P }_{ pq }\sum _{ \sigma  }^{  }{ \left< { \left[ H,{ a }_{ p\sigma  }^{ \dagger  }{ a }_{ q\sigma  } \right]  } \right>  } \\ 
&=\mathcal{ P }_{ pq }\sum _{ uvt }^{  }{ \left[ \left( { uv }|{ tp } \right) { \Gamma  }_{ ut }^{ vq }+\left( { up }|{ tv } \right) { \Gamma  }_{ ut }^{ qv }-\left( { uv }|{ qt } \right) { \Gamma  }_{ up }^{ vt }-\left( { qv }|{ tu } \right) { \Gamma  }_{ pt }^{ vu } \right]  } \\
\end{split}
\end{align}
where $\mathcal{ P }_{ pq }$ is a permutation operator $\mathcal{ P }_{ pq }=1-\left( p\leftrightarrow q \right) $ and the notation for the expectation value means 
\begin{align}
\label{eqn:expectation}
\left< { O } \right> =\left< { \Psi  }|{ \hat { O }  }|{ \Psi  } \right> 
\end{align}

Similarly, the Hessian is
\begin{align}
\label{eqn:hessian:app}
\begin{split}
&{ A }_{ pq,rs }={ \left( \frac { { \partial  }^{ 2 }E\left( \kappa  \right)  }{ \partial { \kappa  }_{ pq }\partial { \kappa  }_{ rs } }  \right)  }_{ \kappa =0 } \\
&=\frac { 1 }{ 2 } \mathcal{ P }_{ pq }\mathcal{ P }_{ rs }\sum _{ \sigma ,\tau  }^{  }{ \left< { \left[ \left[ H,{ a }_{ p\sigma  }^{ \dagger  }{ a }_{ q\sigma  } \right] ,{ a }_{ r\tau  }^{ \dagger  }{ a }_{ s\tau  } \right]  } \right>  } \\
&+\frac { 1 }{ 2 } \mathcal{ P }_{ pq }\mathcal{ P }_{ rs }\sum _{ \sigma ,\tau  }^{  }{ \left< { \left[ \left[ H,{ a }_{ r\sigma  }^{ \dagger  }{ a }_{ s\sigma  } \right] ,{ a }_{ p\tau  }^{ \dagger  }{ a }_{ q\tau  } \right]  } \right>  } \\
\end{split}
\end{align}

We obtain
\begin{align}
&{ A }_{ pq,rs }=\mathcal{ P }_{ pq }\mathcal{ P }_{ rs }\frac { 1 }{ 2 }
\sum _{ uvt }^{  }
{ \delta  }_{ qr }
\bigg(
   \left( { uv }|{ tp } \right) { \Gamma  }_{ ut }^{ vs }+\left( { up }|{ tv } \right) { \Gamma  }_{ ut }^{ sv } \notag \\
& \hspace{40mm}
 + \left( { uv }|{ st } \right) { \Gamma  }_{ up }^{ vt }+\left( { sv }|{ tu } \right) { \Gamma  }_{ pt }^{ vu }
\bigg) \notag \\
&+{ \delta  }_{ ps }\left( \left( { uv }|{ qt } \right) { \Gamma  }_{ ur }^{ vt }+\left( { qv }|{ tu } \right) { \Gamma  }_{ rt }^{ vu }+\left( { uv }|tr \right) { \Gamma  }_{ ut }^{ vq }+\left( { ur }|{ tv } \right) { \Gamma  }_{ ut }^{ qv } \right) \notag \\
&+\sum _{ uv }^{  }{ \left( { up }|{ vr } \right) { \Gamma  }_{ uv }^{ qs }+\left( ur|{ vp } \right) { \Gamma  }_{ uv }^{ sq }+\left( { qv }|{ su } \right) { \Gamma  }_{ pr }^{ vu }+\left( { sv }|{ qu } \right) { \Gamma  }_{ rp }^{ vu } } \notag \\
&-\sum _{ tu }^{  } \left( ut|{ qr } \right) { \Gamma  }_{ up }^{ ts }+\left( qu|{ tr } \right) { \Gamma  }_{ pt }^{ us }+\left( { ur }|{ qt } \right) { \Gamma  }_{ up }^{ st }+\left( { qr }|{ tu } \right) { \Gamma  }_{ pt }^{ su } \notag \\
&+\left( { ut }|{ sp } \right) { \Gamma  }_{ ur }^{ tq }+\left( up|{ st } \right) { \Gamma  }_{ su }^{ tp }+\left( { su }|{ tp } \right) { \Gamma  }_{ rt }^{ uq }+\left( { sp }|{ tu } \right) { \Gamma  }_{ rt }^{ qu } 
\label{eqn:orb_hessian_app}
\end{align}



\bibliographystyle{aip}
\bibliography{vmc_cc.bib}

\begin{thebibliography}{10}

\bibitem{MolElecStruc}
T.~Helgaker, P.~J{\o}gensen, and J.~Olsen,
\newblock {\em Molecular Electronic Structure Theory},
\newblock John Wiley and Sons, Ltd, West Sussex, England, 2000.

\bibitem{Plesset:1934:mp2}
C.~M{\o}ller and M.~S. Plesset,
\newblock Phys. Rev {\bf 46}, 618 (1934).

\bibitem{Parac:2003:failure_tddft}
S.~Grimme and M.~Parac,
\newblock ChemPhysChem {\bf 4}, 292 (2003).

\bibitem{Eric:2016:ScO}
E.~Neuscamman,
\newblock J. Chem. Theory Comput {\bf 12}, 3149 (2016).

\bibitem{Subedi:2008:supercond}
A.~Subedi, L.~Zhang, D.~J. Singh, and M.~H. Du,
\newblock Phys. Rev. B {\bf 78}, 134514 (2008).

\bibitem{NuclearMB}
P.~Ring and P.~Schuck,
\newblock {\em The Nuclear Many-Body Problem},
\newblock Springer, Berlin, 2000.

\bibitem{Scuseria:2011:seniority}
L.~Bytautas, T.~M. Henderson, C.~A. Jim\'{e}nez-Hoyos, J.~K. Ellis, and G.~E.
  Scuseria,
\newblock J. Chem. Phys {\bf 135}, 044119 (2011).

\bibitem{Wilson:1967:doci}
F.~Weinhold and E.~B.~W. Jr.,
\newblock J. Chem. Phys {\bf 46}, 2752 (1967).

\bibitem{Shull:1962:Be}
T.~L. Allen and H.~Shull,
\newblock J. Phys. Chem {\bf 66}, 2281 (1962).

\bibitem{Fogel:1965:Be}
D.~W. Smith and S.~J. Fogel,
\newblock J. Phys. Chem {\bf 43}, S91 (1965).

\bibitem{Barlett:1982:ccsd}
G.~D.~P. III and R.~J. Barlett,
\newblock J. Chem. Phys. {\bf 76}, 1910 (1982).

\bibitem{Martin:1989:ccsd(t)}
K.~Raghavachari, G.~W. Trucks, J.~A. Pople, and M.~Head-Gordon,
\newblock Chem. Phys. Lett. {\bf 157}, 479 (1989).

\bibitem{Scuseria:2014:pccd}
T.~Stein, T.~M. Henderson, and G.~E. Scuseria,
\newblock J. Chem. Phys {\bf 140}, 214113 (2014).

\bibitem{Scuseria:2014:pccd2}
T.~M. Henderson, I.~W. Bulik, T.~Stein, and G.~E. Scuseria,
\newblock J. Chem. Phys {\bf 141}, 244104 (2014).

\bibitem{Scuseria:2015:cc_strong_corr}
I.~W. Bulik, T.~M. Henderson, and G.~E. Scuseria,
\newblock J. Chem. Theory Comput. {\bf 11}, 3171 (2015).

\bibitem{Barlett:2007:cc_rev}
R.~J. Bartlett and M.~Musial,
\newblock Rev. Mod. Phys {\bf 79}, 291 (2007).

\bibitem{barlett:2012:mrcc}
D.~I. Lyakh, M.~Muslal, V.~F. Lotrlch, and R.~J. Barlett,
\newblock Chem. Rev {\bf 112}, 182 (2012).

\bibitem{Knowles:2010:vcc}
B.~Cooper and P.~J. Knowles,
\newblock J. Chem. Phys {\bf 133}, 234102 (2010).

\bibitem{Scuseria:2016:ccd0}
J.~A. Gomez, T.~M. Henderson, and G.~E. Scuseria,
\newblock J. Chem. Phys {\bf 144}, 244117 (2016).

\bibitem{Martin:2014:ccvb}
D.~W. Small, K.~V. Lawler, and M.~Head-Gordon,
\newblock J. Chem. Theory Comput {\bf 10}, 2027 (2014).

\bibitem{Ayers:2013:ap1rog}
P.~A. Limacher et~al.,
\newblock J. Chem. Theory Comput {\bf 9}, 1394 (2013).

\bibitem{Knowles:2012:quasi-vcc}
J.~B. Robinson and P.~J. Knowles,
\newblock J. Chem. Phys {\bf 136}, 054114 (2012).

\bibitem{Umrigar:2015:qmc}
C.~J. Umrigar,
\newblock J. Chem. Phys. {\bf 143}, 164105 (2015).

\bibitem{Sorella:2003:jagp}
M.~Casula and S.~Sorella,
\newblock J. Chem. Phys. {\bf 119}, 6500 (2003).

\bibitem{Umrigar:2005:lm}
C.~J. Umrigar and C.~Filippi,
\newblock Phys. Rev. Lett {\bf 94}, 150201 (2005).

\bibitem{Nightingale:2001:linear_method}
M.~P. Nightingale and V.~Melik-Alaverdian,
\newblock Phys. Rev. Lett. {\bf 87}, 043401 (2001).

\bibitem{UmrTouFilSorHen-PRL-07}
C.~J. Umrigar, J.~Toulouse, C.~Filippi, S.~Sorella, and R.~G. Hennig,
\newblock Phys. Rev. Lett. {\bf 98}, 110201 (2007).

\bibitem{Pyscf_brief}
Pyscf, a quantum chemistry package written in python,
\newblock see http://chemists.princeton.edu/chan/software/pyscf.

\bibitem{MOLPRO_brief}
H.-J. Werner et~al.,
\newblock \uppercase{MOLPRO}, version 2012.1, a package of ab initio programs,
\newblock see http://www.molpro.net.

\bibitem{Psi4:2012}
J.~M. Turney et~al.,
\newblock WIREs Comput. Mol. Sci. {\bf 2}, 556 (2012).

\bibitem{person_comm}
P.~A. Limacher,
\newblock personal communication, 2016.

\bibitem{cc-pvdz}
T.~Dunning,
\newblock J. Chem. Phys {\bf 90}, 1007 (1989).

\bibitem{6-31g}
W.~J. Hehre, R.~Ditchfield, and J.~A. Pople,
\newblock J. Chem. Phys {\bf 56}, 2257 (1972).

\end{thebibliography}

\end{document}